\documentclass[draftcls,12pt,onecolumn]{IEEEtran}

\usepackage{graphicx}
\ifCLASSINFOpdf
\else
\fi
\usepackage[cmex10]{amsmath}
\usepackage{algorithm}
\usepackage{algorithmicx}
\usepackage{algpseudocode}

\begin{document}
%
\title{Simultaneous Wireless Information and Power Transfer for Two-hop OFDM Relay System}
%
%
%

\author{Xiaofei Di,
        Ke Xiong
        and Zhengding Qiu
        \thanks{The authors are with school of computer and information technology, Beijing Jiaotong University, Beijing, 100044, P. R. China. (e-mail: \{09112084, kxiong, zdqiu\}@bjtu.edu.cn).}
        }

\markboth{}%
{Shell \MakeLowercase{\textit{et al.}}: Bare Demo of IEEEtran.cls for Journals}
%



\maketitle

\begin{abstract}
This paper investigates the simultaneous wireless information and power transfer (SWIPT) for two-hop orthogonal frequency division multiplexing (OFDM) decode-and-forward (DF) relay communication system, where a relay harvests energy from radio frequency signals transmitted by the source and then uses the harvested energy to assist the information transmission from the source to its destination. The power splitting receiver is considered at the relay, which splits the received signal into two power streams to perform information decoding (ID) and energy harvesting (EH) respectively. For better understanding the behavior and exploring the performance limit of such a system, resource allocation is studied to maximize the total achievable transmission rate. An optimization problem, which jointly takes into account the power allocation, the subcarrier pairing and the power splitting, is formulated. Due to its non-convexity, a resource allocation policy with low complexity based on separation principle is designed. Simulation results show that the system performance can be significantly improved by using our proposed policy. Moreover, the system performance behavior to the relay position is also discussed, and results show that in the two-hop OFDM system with EH relay, the relay should be deployed near the source, while in that with conventional non-EH relay, it should be deployed at the middle between the source and the destination.

\end{abstract}


%
\IEEEpeerreviewmaketitle

\section{Introduction}
%
%
%
%

%
%

Recently, relay communication has been widely investigated to improve the system capacity, reduce energy consumption and extend communication coverage \cite{1}. In relay communication system, relay nodes can employ some relaying protocols to assist to transmit information from a source node to its destination node. Among existing relaying protocols, amplify-and-forward (AF) and decode-and-forward (DF) are the most popular protocols \cite{2}, \cite{Kramer}.

Meanwhile, orthogonal frequency-division multiplexing (OFDM) has been adopted as the air interface in broadband wireless networks. By converting a frequency-selective wideband channel into a set of orthogonal narrowband frequency flat subcarrier channels, OFDM not only eliminates the intersymbol interference effectively, but also provides good design flexibility \cite{Yu}.

The combination of relaying and OFDM techniques is believed to be further able to enhance the system performance \cite{Wang}-\cite{MTao}. For OFDM relay system, resource allocation is an important issue and thus needs to be appropriately designed in order to acquire the better system performance. Firstly, the incoming and outgoing subcarriers need to be carefully paired at the relay according to the channel conditions over the two hops. This is well-known as subcarrier pairing. Secondly, allocating power properly among subcarriers can also improve system performance significantly.

For some applications of OFDM relay system, such as wireless sensor networks (WSN) or wireless body area networks (WBAN), connecting the source or the relay to the power grid may be impossible. Although batteries can be deployed to solve this problem, the limited capacity and high transmit power may lead to quick depletion of the batteries. Therefore, the batteries are requested to be replaced or recharged frequently.

Recently, an alternative solution is proposed, i.e., to deploy energy harvesting (EH) technique in the system, where EH nodes can harvest energy from their surrounding environment to maintain their operation. The EH technique mainly includes two different ways. One way is to harvest energy from solar, thermoelectric, and some other physical phenomena \cite{3}. But it is heavily dependent of surrounding environment and cannot supply continuous energy source, so it is hard to support EH nodes' and the system's steady work.

Another way is to harvest energy from radio frequency (RF) signals from other nodes which are supplied with stable energy source, including connecting power gird or having high capacity batteries. This way is also called as simultaneous wireless information and power transfer (SWIPT) \cite{7}-\cite{4}, since RF signals carry both energy and information simultaneously, and it is regarded as a promising option to ensure a long system lifetime without the requirement for periodic battery replacements and without the dependence upon surrounding environment.

For SWIPT, an ideal receiver was assumed so that information decoding (ID) and energy harvesting can be performed simultaneously from the same RF signal in the primary works (see e.g. \cite{5}, \cite{6}). However, this assumption is considered to be impractical for real wireless systems, since circuits for harvesting energy from RF signals cannot be used to decode the carried information directly. So two practical SWIPT architectures, namely, time switching (TS) and power splitting (PS), were proposed in \cite{7}. When TS is applied at the receiver, the received signal is either processed by an energy receiver for EH or processed by an information receiver for ID from the perspective of time. When PS is applied at the receiver, the received signal is split into two power streams by a power splitter, with one stream to the energy receiver and the other one to the information receiver. Later, these two practical architectures were further applied to two-hop relay system \cite{8} and one-hop OFDM system \cite{EHofdm1}, \cite{EHofdm2}. In \cite{8}, the outage performance for two-hop relay system with TS and PS deployed respectively at the relay node was investigated.  In \cite{EHofdm1} and \cite{EHofdm2}, energy efficiency and weighted sum-rate metrics were optimized for multiuser OFDM systems where PS or TS is employed at mobile users. 

As far as we know, the SWIPT technique has not been studied in two-hop OFDM relay system. Since SWIPT technique brings a new degree of freedom for OFDM relay system's design, it makes the system's design different from the conventional OFDM relay system, which motivates us to discuss the performance of such a system. 

In this paper, we consider SWIPT for a three-node two-hop OFDM relay system, where a source communicates with its destination via a half-duplex DF relay. It is assumed that the source has fixed energy supply while the relay has no fixed energy supply. The relay is an EH node and needs to harvest energy from RF signal transmitted by the source. Then with the harvested energy, it can help the information transmission from the source to its destination. PS is applied at the relay. In order to explore the performance limit of such a system, we investigate the resource allocation for it and then formulate a non-convex optimization problem to maximize the total achievable transmission rate of the system, in which subcarrier pairing, power allocation and power splitting are jointly optimized. 

The main contributions of this paper are summarized as follows. We design the separation principle based resource allocation policy to solve the optimization problem. For the subcarrier pairing, the incoming and outgoing subcarriers are matched at the relay according to two hops¡¯ channel gains, and for the power splitting and the power allocation, the explicit solutions are obtained. Our results reveal that the presented system's performance obtains great improvement by using our proposed policy. The performance of the system with EH relay is also compared with that with conventional non-EH relay. Results show that due to energy loss, the system with EH relay may loss some performance compared to that with conventional relay. But they show very different behavior w.r.t the relay position. That is,  in the two-hop OFDM system with EH relay, the relay should be deployed near the source, while in that with conventional relay, it should be deployed at the middle between the source and the destination.

This paper is organized as follows. In Section II the system model is presented. In Section III, we formulate an resource allocation optimization problem and then design a resource allocation in Section IV. Simulation results are shown in Section V for performance evaluations and comparisons. Finally, Section VI summarizes this paper.

\section{System Model}

Consider a two-hop relay OFDM system, as shown in Fig. \ref{systemmodel}, which consists of one source, one destination and one relay. The source wants to transmit information to the destination with the assistance of the relay, and it is assumed that there is no the direct link from the the source to its destination. The relay operates in half-duplex mode, which means it cannot simultaneously transmit and receive signals. Decode-and-forward (DF) relaying protocol is employed at the relay. The source is with fixed power supply, and its available power is denoted by $P_{\max}$. The relay is an energy harvesting (EH) node, which has to harvest the energy from RF signal transmitted by the source to help the data transmission from the the source to its destination. 


For such a system, the two-hop EH transmission is on a time-frame basis with each frame consisting of multiple OFDM symbols. Each frame is of length $T$, which is further divided into two equal time slots. In the first time slot, the source transmits OFDM symbols to the relay. Thus, the received signal at the relay over subcarrier $i$ can be given by
\[{y_{r,i}} = \sqrt {{P_{s,i}}} {h_{i}}{x_i} + {z_{r,i}}, \forall i\in\{1,...,N\},\]
where $N$ is the total subcarrier number of the system, $x_i$, $P_{s,i}$ and $h_{i}$ denote the transmitted data symbol, the transmit power at the source and the fading coefficient over subcarrier $i$, respectively. $z_{r,i}$ represents the additive white Gaussian noise (AWGN) with zero mean and variance $\sigma _{r,a}^2$ from the antenna over subcarrier $i$ at the relay. The relay adopts PS scheme to split the received signal into two power streams to perform EH and information decoding (ID), respectively. The power splitting ratios $\rho^I_i$ and $\rho^E_i$ represent the fraction of the received signal power used for ID and EH, respectively, which satisfy the constraints of
\begin{equation}
\rho^I_i+\rho^E_i=1,\ \ \rho_i^I\geq0,\ \ \rho_i^E\geq0, \forall i.
\label{C1}
\end{equation}

Therefore, the harvested energy over subcarrier $i\in\{1,...,N\}$ at the relay can be given by
$E_{i} = \tfrac{T}{2}\eta {\rho ^E_i}{\left| {{h_i}} \right|^2}P_{s,i},$
where $\eta$ is a constant, denoting the energy harvesting efficiency and satisfying $0\leq\eta\leq1$.

In the second time slot, using the power stream for ID, the relay decodes the received symbols and then forward re-encoded symbols to the destination by using the harvested energy. We consider such an \emph{energy cooperation} strategy, in which the energy harvested over subcarrier $i$ of the first hop is only used to transmit the received information over its corresponding subcarrier $j$ of the second hop. That is, with subcarrier pairing, the information received over subcarrier $i$ is forwarded over subcarrier $j$, as shown in Fig. \ref{systemmodel}. Thus the received signal over subcarrier $j$ at the destination in this time slot can be expressed as
$${y_{d,j}} = \sqrt {{P_{r,j}}} {g_j}{x_i} + {z_{d,j}}, \forall j\in\{1,...,N\},$$
where $g_{j}$ is fading coefficient over subcarrier $j$ at the destination, $z_{d,j}$ is AWGN with zero mean and variance $\sigma _{d,a}^2$ from the antenna over subcarrier $j$ at the destination. $P_{r,j}$ is the available power over subcarrier $j$ at the relay in the second time slot. Thus, according to our proposed energy cooperation strategy, $P_{r,j}$ can be given by
\begin{equation}
{P_{r,j}} = \frac{E_{i}}{{T/2}} = \eta {\rho ^E_i}{\left| {{h_i}} \right|^2}P_{s,i}.
\label{Pri}
\end{equation}


\section{Optimization Problem Formulation}

In order to explore the performance limit of the presented EH OFDM relay system, we investigate the resource allocation for it. To do so, we formulate an optimization problem to maximize the total achievable transmission rate of the system.

As is known, the achievable transmission rate between the source and its destination for DF relay system over a subcarrier pair ($i,j$) can be given by
\begin{equation}
R_{i,j} = \frac{1}{2}\min \left\{\log_2 \big(1 + \frac{{\rho _i^I{\left| {{h_i}} \right|^2}}{P_{s,i}}}{{\rho _i^I\sigma _{r,a}^2 + \sigma _{r,b}^2}}\big),\log_2 \big(1 + \frac{{\left| {{g_j}} \right|^2}{P_{r,j}}}{{\sigma _d^2}}\big)\right\},
\label{Rij1}
\end{equation}
where $\sigma _d^2=\sigma _{d,a}^2+\sigma _{d,b}^2$ is total noise power over each subcarrier at the destination. $\sigma _{r,b}^2$ and $\sigma _{d,b}^2$ are the signal processing noise power over each subcarrier at the relay and the destination, respectively \cite{EHofdm1}. The first term (i.e. $\log_2 (1 + \frac{{\rho _i^I{\left| {{h_i}} \right|^2}}{P_{s,i}}}{{\rho _i^I\sigma _{r,a}^2 + \sigma _{r,b}^2}})$ in (\ref{Rij1}) is the mutual information from the source to the relay over subcarrier $i$, and the second term (i.e. $\log_2 (1 + \frac{{\left| {{g_j}} \right|^2}{P_{r,j}}}{{\sigma _d^2}})$)  in (\ref{Rij1}) is the mutual information from the relay to the destination over subcarrier $j$. The pre-log factor $\frac{1}{2}$ is due to two time slots in each frame.

Substituting (\ref{Pri}) into (\ref{Rij1}) and denoting $P_{s,i}$ as $P_{i}$ for simplicity, then we can write (\ref{Rij1}) as
\begin{equation}
R_{i,j} =
\frac{1}{2}\min \left\{\log_2 \big(1 + \frac{{\rho _i^I{\left| {{h_i}} \right|^2}{P_i}}}{{\rho _i^I\sigma _{r,a}^2 + \sigma _{r,b}^2}}\big),\log_2 \big(1 + \frac{{{\eta}\rho _i^E{\left| {{h_i}} \right|^2}{\left| {{g_j}} \right|^2}{P_i}}}{{\sigma _d^2}}\big)\right\}.
\label{eq1}
\end{equation}

Hence, the total achievable transmission rate of the system can be defined as
\[R(\mathcal{P},\mathcal{S},\boldsymbol{\rho} ) = \sum\nolimits_{i = 1}^N {\sum\nolimits_{j = 1}^N {{s_{i,j}}{R_{i,j}}} }, \]
where $\mathcal{P}=\{P_i,\forall i\}$ represents the power allocation, which satisfies that
\begin{equation}
  \sum\nolimits_{i = 1}^N {{P_i}}  = {P_{\max }},\ \ P_i\geq0, \forall i.
\label{C2}
\end{equation}
$\boldsymbol{\rho}=\{\rho_i^I, \rho_i^E, \forall i\}$ is the power splitting policy and satisfies the constraints (\ref{C1}).  $\mathcal{S}=\{s_{i,j}, \forall i,j\}$ is the subcarrier pairing policy, which means if incoming subcarrier $i$ and outgoing subcarrier $j$ are matched at the relay, $s_{i,j}=1$, otherwise   $s_{i,j}=0$. Thus, it satisfies
\begin{equation}
s_{i,j} \in \{ 0,1\} ,\forall i,j,
\ \ \sum\nolimits_{j = 1}^N {{s_{i,j}}}  \le 1,\forall i,
\ \  \sum\nolimits_{i = 1}^N {{s_{i,j}}}  \le 1,\forall j.
\label{C3}
\end{equation}

As a result, the optimization problem can be formulated as ($\textbf{P1}$):
\[\begin{array}{l l}
&\mathop {\max }\limits_{\mathcal{P},\mathcal{S},\boldsymbol{\rho} } R(\mathcal{P},\mathcal{S},\boldsymbol{\rho} ) \\
&{\rm{s}}{\rm{.t}}{\rm{.  }} \ \ (\ref{C1}),(\ref{C2}),(\ref{C3}).
\end{array}\]



\section{Resource Allocation Design}

As $\textbf{P1}$ is a mixed integer programming problem, which is non-convex and difficult to solve, in this section, we propose a resource allocation policy on the basis of a separation principle to solve the problem and prove the optimality of the proposed policy.

\subsection{The Proposed Resource Allocation}
The proposed resource allocation policy can be described as Algorithm \ref{Algorithm1}. Although the separation design is adopted in Algorithm \ref{Algorithm1}, it still can be demonstrated to achieve the global optimal solution of $\textbf{P1}$. In the following, we shall first describe the detailed  process of each step of Algorithm \ref{Algorithm1} and then prove the global optimum of it in subsection \ref{SEC:B}.

\begin{algorithm}
\caption{: Resource Allocation Algorithm}
\begin{algorithmic}
\State  1) Find the optimal subcarrier pairing $\mathcal{S}^*$ based on the channel gains of the two hops.
\State  2) Calculate the optimal power splitting $\boldsymbol{\rho}^*$ with the obtained optimal subcarrier pairing $\mathcal{S}^*$.
\State  3) Calculate the  optimal power allocation $\mathcal{P}^*$  with the obtained optimal subcarrier pairing $\mathcal{S}^*$ and optimal power splitting $\boldsymbol{\rho}^*$.
\end{algorithmic}
\label{Algorithm1}
\end{algorithm}

\subsubsection{The Optimal Subcarrier Pairing $\mathcal{S}^*$}
The proposed subcarrier pairing scheme is only dependent on the channel gains of two hops. First, the incoming subcarriers are sorted from most to least according to the channel gains $\left| {{h_i}} \right|^2$ from the source to the relay. Then the the outgoing subcarriers are also sorted from most to least according to the channel gains $\left| {{g_j}} \right|^2$ from the relay to the destination. Finally, the \emph{k}-th incoming subcarrier is paired with the \emph{k}-th outgoing subcarrier for all $k\in \{1,...,N\}$. We refer to this scheme as the sorted pairing in the sequel.

\subsubsection{The Optimal Power Splitting $\boldsymbol{\rho}^*$ with the obtained $\mathcal{S}^*$}

 First, it is easily found that the problem can be decomposed into $N$ subproblems due to the independence of each subcarrier pair. For any subcarrier pair $(i,j)$, the subproblem can be expressed as ($\textbf{P2}$):
\[
\begin{array}{l l}
&\mathop {\max }\limits_{\rho _i^I,\rho _i^E}{R_{i,j}} \\
&{\rm{s}}{\rm{.t}}{\rm{.  }}
\ \  \rho _i^I + \rho _i^E = 1, \rho _i^I \ge 0, \rho _i^E \ge 0.
\end{array}
\]

Since the subcarrier pair is fixed, we can drop the indexes $i,j$ in this subsection. To simplify the expressions, let $a={{\left| {{h_i}} \right|^2}}$, $b=\frac{\eta {\left| {g_j} \right|^2}}{\sigma^2_d}$. Thus the achievable transmission rate in (\ref{eq1}) over a fixed subcarrier pair can be expressed as
\begin{equation}
R = \frac{1}{2}\min \left\{\log_2 \big(1 + \frac{a\rho^IP}{\rho^I\sigma_{r,a}^2+\sigma_{r,b}^2}\big), \log_2 (1 + ab\rho^EP)\right\}.
\label{eq2}
\end{equation}


It is easy to find that in (\ref{eq2}) the first term (i.e. $\log_2 (1 + \frac{a\rho^IP}{\rho^I\sigma_{r,a}^2+\sigma_{r,b}^2})$ is a monotonically increasing function of $\rho^I$, and the second term (i.e. $\log_2 (1 + ab\rho^EP)$ is a monotonically decreasing function of $\rho^I$, so to obtain the optimal solution, the two terms should be equal. Meanwhile, using $\rho^I+\rho^E=1$, the optimal power splitting ratio $\rho^I$ can be computed, according to
$$
b\sigma_{r,a}^2(\rho^I)^{2}+(1-b\sigma_{r,a}^2+b\sigma_{r,b}^2)\rho^I-b\sigma_{r,b}^2=0.
$$


This is a quadratic equation with the variable $\rho^I$, and we can easily obtain its two roots as
\[
\rho^I=\tfrac{-(1-b\sigma_{r,a}^2+b\sigma_{r,b}^2)\pm\sqrt{(1-b\sigma_{r,a}^2+b\sigma_{r,b}^2)^2+4b^2\sigma_{r,a}^2\sigma_{r,b}^2}}{2b\sigma_{r,a}^2}.
\]

Howerer, $\rho^I$, which satisfies the constraints in $\textbf{P2}$ , can only be considered as the optimal solution. Since $\rho^I+\rho^E=1,\rho^I,\rho^E\geq0$, we have that $0\leq\rho^I,\rho^E\leq1$.
It is easy to prove that the one of the two roots $\rho^I=\tfrac{-(1-b\sigma_{r,a}^2+b\sigma_{r,b}^2)-\sqrt{(1-b\sigma_{r,a}^2+b\sigma_{r,b}^2)^2+4b^2\sigma_{r,a}^2\sigma_{r,b}^2}}{2b\sigma_{r,a}^2}$ is always less than 0, so it is discarded. For the other one, we can prove that it satisfies the above constraint.

For this, we consider two extreme cases. First for the case that $\rho^I=0, \rho^E=1$, we have the first term in (\ref{eq2}) is equal to 0, the second term is larger than 0. And then for the case that $\rho^I=1, \rho^E=0$, we have the second term in (\ref{eq2}) is equal to 0, the first term is larger than 0. So the intersection point of the corresponding curves of the two terms certainly locates in the interval [0,1], as shown in Fig. \ref{curves}. So the other root satisfies the aforementioned condition. Consequently, the optimal power splitting ratios can be given by
\begin{equation}
\begin{array}{l}
\rho^{I*}=\frac{-(1-b\sigma_{r,a}^2+b\sigma_{r,b}^2)+\sqrt{(1-b\sigma_{r,a}^2+b\sigma_{r,b}^2)^2+4b^2\sigma_{r,a}^2\sigma_{r,b}^2}}{2b\sigma_{r,a}^2},\\
\rho^{E*}=1-\rho^{I*}.
\end{array}
\label{optrho}
\end{equation}

\subsubsection{The Optimal Power Allocation $\mathcal{P}^*$ with the obtained $\mathcal{S}^*$ and $\boldsymbol{\rho}^*$}

After the optimal power splitting policy is designed, we obtain the optimal power splitting ratios $\rho^{I*}_{i}$ and $\rho^{E*}_{i}$. As the optimal power splitting ratios are related to the channel gain of the second hop from (\ref{optrho}), $\rho^{I*}_{i}, \rho^{E*}_{i}$ is represented as $ \rho^{I*}_{i,j}, \rho^{E*}_{i,j}$ for given subcarrier pair $(i,j)$. Since the two terms are equal in (\ref{eq1}) for optimal $ \rho^{I*}_{i,j}, \rho^{E*}_{i,j}$, (\ref{eq1}) can be transformed as
\begin{equation}
R_{i,j} = \frac{1}{2}\log_2 \big(1 + \frac{{\left| {{h_i}} \right|^2}\rho^{I*}_{i,j}P_i}{\rho^{I*}_{i,j}\sigma_{r,a}^2+\sigma_{r,b}^2}\big).
\label{eq5}
\end{equation}

We denote $\frac{{\left| {{h_i}} \right|^2}\rho^{I*}_{i,j}}{\rho^{I*}_{i,j}\sigma_{r,a}^2+\sigma_{r,b}^2}$ as $\gamma_{i,j}$, and then (\ref{eq5}) becomes
\begin{equation}
R_{i,j} = \frac{1}{2}\log_2 (1 + \gamma_{i,j}P_i).
\label{rijf}
\end{equation}

So the power allocation problem can be formulated as ($\textbf{P3}$):
\[\begin{array}{l l}
&\mathop {\max }\limits_{\mathcal{P} } \sum\limits_{(i,j) \in \mathcal{SP}} {R_{i,j}} \\
&{\rm{s}}{\rm{.t}}{\rm{.  }} \ \ (\ref{C2}),
\end{array}\]
where $\mathcal{SP}=\{(i,j)| s_{i,j}=1, \forall i,j\}$ represents the set of subcarrier pairs. From the optimal power splitting expressions (\ref{optrho}), it can be easily found that the optimal power splitting ratios are not related to power allocation and thus $\gamma_{i,j}$ is also not related to the power allocation. So this problem is a classical water-filling problem, and we can obtain its optimal solution as
\[
P_i^*=\left[\frac{1}{\nu\ln2}-\frac{1}{\gamma_{i,j}}\right]^+, \forall i,
\]
where $[x]^+=\max\{0,x\}$, and $\nu$ is Lagrangian multiplier which can be solved using the equation
$\sum\nolimits_{i=1}^N[1/(\nu\ln2)-{1}/{\gamma_{i,j}}]^+=P_{\max}.$

\subsection{The Global Optimum of Our Proposed Resource Allocation}\label{SEC:B}

In subsection A, the proposed resource allocation based on a separation strategy is presented. In this subsection, we  shall prove that our proposed policy in Algorithm 1 achieves the global optimal solution of the problem \textbf{P1}. For this, we first give the following lemma.

\newtheorem{lemma}{Lemma}
\begin{lemma}
The sorted subcarrier pairing scheme in Algorithm \ref{Algorithm1} is globally optimal for the problem \textbf{P1}.
\label{T1}
\end{lemma}

\begin{IEEEproof}
To prove this lemma, two-subcarrier case is first considered and proved. Then it is extended to general multi-subcarrier case. See Appendix A for details.
\end{IEEEproof}

Next, using this lemma, we discuss the global optimum of the proposed resource allocation policy and the result is given by the following theorem.

\newtheorem{theorem}{Theorem}
\begin{theorem}
The resource allocation policy in Algorithm 1 achieves the global optimal solution of the problem \textbf{P1}.
\label{T2}
\end{theorem}

\begin{IEEEproof}
To prove that the resource allocation policy in Algorithm 1 can achieve the global optimal solution of the problem \textbf{P1}, we only have to prove that each step of Algorithm 1 maintains the global optimum. From Lemma \ref{T1}, we have known that the sorted pairing scheme in the first step of Algorithm 1 gives the globally optimal subcarrier pairing policy. The scheme is only related to channel gains, which does not require the knowledge of the optimal power splitting and power allocation. According to the derivation process in the second step of Algorithm 1, the obtained power splitting is optimal under the given optimal subcarrier pairing, and it does not require  the knowledge of optimal power allocation. Then in the third step of Algorithm \ref{Algorithm1}, the obtained power allocation is optimal under the given optimal subcarrier pairing and optimal power splitting. Since each step maintains the global optimum, Theorem 1 is proved.
\end{IEEEproof}

Then we analyze the computational complexity of the proposed resource allocation policy. The complexity of the sorted subcarrier pairing in the first step depends on the adopted sorting method, which is $O(N\log N)$ for the best sorting algorithm. The complexity of power splitting is $O(N)$, and the complexity of water-filling power allocation is also $O(N)$ under the condition all $\gamma_{i,j}$ are sorted. Thus, the total computational complexity is $O(N\log N)$, and our proposed policy is low complexity algorithm.

\section{Simulation Results}

In this section, we first demonstrate the performance of the proposed resource allocation policy and then compare the performances of the OFDM system with EH relay and that with conventional non-EH relay via simulation results. In our simulations, it is assumed that the noise powers  $\sigma_d^2, \sigma_{r,a}^2, \sigma_{r,b}^2$  are 1dBm. The energy harvesting efficiency is set to $\eta=1$. We set the number of subcarrier to be $N=4$. It is assumed that all the three nodes are located on a line. The distance between source and destination is denoted by $d_0$ as a reference distance, and the location of the relay is normalized as $\frac{d_r}{d_0}$, where $d_r$ is the distance between the source and the relay. Channel coefficients $h_i$ (or $g_j$) are picked from a Rayleigh fading channel with the distribution \cite{9} as
$
h_i=\rm{{\cal C}{\cal N}}\big(0,\frac{1}{\emph{L}(1+\emph{d})^\alpha}\big),
$
where the path loss exponent $\alpha$ is set to 3, $d$ is the distance between the source and the relay (or between the relay and the destination for $g_j$), and the number of taps is set to $L=4$.

\subsection{Effect of Resource Allocation Policy and Relay Location}


In this subsection, we discuss the effects of the resource allocation policy and the location of the relay on the achievable transmission rate of the system. For comparisons, three suboptimal policies are also simulated, i.e., 1) Optimal power allocation without subcarrier pairing; 2) Uniform power allocation with subcarrier pairing; 3) Uniform power allocation without subcarrier pairing. In order to fairly compare the three policies with our proposed policy, the optimal power splitting is employed for each of them. In Fig. \ref{Fig3}, we plot the average achievable rate versus the total power $P_{\max}$. It is easily observed that the proposed policy with power allocation and subcarrier pairing is superior to the other policies.

Fig. \ref{Fig4}. presents the average achievable rate versus the relay's normalized distance. It shows that when the relay node is located near the source node or the destination node, the achievable transmission rate is higher, and further the optimal performance is achieved when the relay is located close to the source.

\subsection{Performance Comparisons with Conventional non-EH relay system}
In this subsection, we compare the performances of the OFDM system with EH relay and that with conventional non-EH relay. It is assumed that the two systems have the identical total available energy in one time-frame. For the system with non-EH relay, let $\tilde{P}_{s,i}$ be the power over subcarrier $i$ at the source, and $\tilde{P}_{r,i}$ be the power over subcarrier $i$ at the relay. Thus, the consumed energy in one frame is
$\tilde{E}=\frac{T}{2}\sum\nolimits_{i=1}^N{\tilde{P}_{s,i}}+\frac{T}{2}\sum\nolimits_{i=1}^N{\tilde{P}_{r,i}}$.
And for the system with EH relay the consumed energy in one frame is $E=\frac{T}{2}\sum\nolimits_{i=1}^N{P_{s,i}}=\frac{T}{2}P_{\max}$. To guarantee the comparison fairness, $\tilde{E}$ should be equal to $E$. So for the system with non-EH relay, it should satisfy the total power constraint
$
\sum\nolimits_{i=1}^N{\tilde{P}_{s,i}}+\sum\nolimits_{i=1}^N{\tilde{P}_{r,i}}=P_{\max}.
$

With this constraint, we performed the optimal resource allocation for the conventional OFDM relay system by using the method proposed in \cite{10}.



The two systems' performances are compared in Fig. \ref{Fig5}. and Fig. \ref{Fig6}. Fig. \ref{Fig5}. compares  their average achievable transmission rates against total power $P_{\max}$. One can observe that the OFDM system with EH relay always get lower achievable transmission rate than that with conventional relay. Moreover, it is also seen that when $P_{\max}$ is relatively low, the gap between the two systems is relatively small, and with the increment of $P_{\max}$, such gap gradually becomes larger.

{Fig. \ref{Fig6}. plots their average achievable transmission rates against the relay's position. It can be easily found the closer the relay is placed to the source, the better performance the system can get. Such an observation is very different from conventional non-EH relay OFDM system, where the maximum achievable rate is achieved only when the relay node is located at the middle point of the source node and its destination node. Moreover, both figures show that the performance of the system with conventional relay is superior to that of the system with EH relay. This is because the relay only harvests energy from fading RF signal transmitted by the source, which way leads to energy loss. That is why the EH relay should be deployed closer to the source.

\section{Conclusion}
In this paper, we investigated the SWIPT for the two-hop OFDM relay communication system. For exploring the presented system's performance limit, we proposed a separation principle based resource allocation policy to maximize the total achievable transmission rate of the system. In simulations, we demonstrated the effect of our proposed resource allocation policy on the achievable transmission rate of the system and results showed that the significant performance gains can be obtained by using our proposed policy. At last, the performance of the OFDM system with EH relay and that with conventional non-EH relay was compared, and simulation results showed that in the system with EH relay, the relay should be deployed near the source, while in that with conventional relay, it should be deployed at the middle between the source and the destination.


%

\appendices
\section{PROOF OF THEOREM 1}

\emph{a)} Two-subcarrier case ($N=2$): Firstly, it is assumed that two hops' channel gains satisfy $\left| {h_1} \right|^2>\left| {h_2} \right|^2, \left| {g_1} \right|^2>\left| {g_2} \right|^2$. Observing Algorithm 1, we know the second step and the third step can be applied to any given subcarrier pairing policy in fact, although they are derived from the optimal subcarrier pairing policy. So from (\ref{rijf}), the achievable transmission rate of the system using sorted subcarrier pairs (1,1) and (2,2) can be expressed as
$
R_{\textrm{sort}}= \frac{1}{2}\log_2 (1 + \gamma_{1,1}P_1^*)+ \frac{1}{2}\log_2 (1 + \gamma_{2,2}P_2^*),
$
and the achievable rate using non-sorted subcarrier pairs (1,2) and (2,1) can be expressed as
$
R_{\textrm{nonsort}}= \frac{1}{2}\log_2 (1 + \gamma_{1,2}P_1^{'*})+ \frac{1}{2}\log_2 (1 + \gamma_{2,1}P_2^{'*}),
$
where $P_1^{'*}, P_2^{'*}$ are the optimal powers for non-sorted pairing scheme. .

To prove Lemma 1, we need to prove that $R_{\textrm{sort}}>R_{\textrm{nonsort}}$, that is,
\begin{equation}
(1 + \gamma_{1,1}P_1^*)(1 + \gamma_{2,2}P_2^*)>(1 + \gamma_{1,2}P_1^{'*})(1 + \gamma_{2,1}P_2^{'*}).
\label{prove1}
\end{equation}

We define a new function $f(\rho^{I*}_{i,j})=\frac{\rho^{I*}_{i,j}}{\rho^{I*}_{i,j}\sigma_{r,a}^2+\sigma_{r,b}^2}$, so $\gamma_{i,j}=\left| {h_i} \right|^2f(\rho^{I*}_{i,j})$.  For further simplifying the expressions, let $H_i=\left| {h_i} \right|^2$ and $G_j=f(\rho^{I*}_{i,j})$. Note that from (\ref{optrho}), we can observe that for given noise power, the optimal power splitting ratio $\rho^{I*}_{i,j}$ is only related to $b$, that is, to the channel gain $\left| {g_j} \right|^2$ of the second hop since $b={\eta {\left| {g_j} \right|^2}}/{\sigma^2_d}$. Thus, for $G_j$, we only reserve the subscript $j$. Seeing that $\gamma_{i,j}=H_iG_j$, (\ref{prove1}) is equivalent to
\begin{equation}
\begin{array}{l}
  (H_1G_1P_1^*+H_2G_2P_2^*+H_1G_1H_2G_2P_1^*P_2^*)-(H_1G_2P_1^{'*}+H_2G_1P_2^{'*}+H_1G_2H_2G_1P_1^{'*}P_2^{'*}) >0.
\end{array}
\label{prove2}
\end{equation}

Secondly, we can prove that $G_1>G_2$ for the assumption $\left| {g_1} \right|^2>\left| {g_2} \right|^2$. From (\ref{optrho}), the derivative of $\rho^{I*}_{i,j}$ with respect to $b$ can be computed as
\[
(\rho^{I*}_{i,j})'=\tfrac{1}{2\sigma_{r,a}^2b^2}\big(1-\tfrac{\frac{1}{b}-(\sigma_{r,a}^2-\sigma_{r,b}^2)}{\sqrt{(\frac{1}{b}-(\sigma_{r,a}^2-\sigma_{r,b}^2))^2+4\sigma_{r,a}^2\sigma_{r,b}^2}}\big).
\]

One can easily find that $(\rho^{I*}_{i,j})'>0$, that is to say, $\rho^{I*}_{i,j}$ is a monotonically increasing function of $b$. Meanwhile we know $b={\eta {\left| {g_j} \right|^2}}/{\sigma^2_d}$, so with the increment of $\left| {g_j} \right|^2$, $\rho^{I*}_{i,j}$ increases. The increment of $\rho^{I*}_{i,j}$ will further result in the increment of $f(\rho^{I*}_{i,j})$. Thus, according to the assumption $\left| {g_1} \right|^2>\left| {g_2} \right|^2$, we have $f(\rho^{I*}_{i,1})>f(\rho^{I*}_{i,2})$, that is, $G_1>G_2$.

Thirdly, for the two-subcarrier case $N=2$, the explicit solutions of optimal power allocation can be obtained. When only equality constraint in (\ref{C2}) is considered and inequality constraints are ignored, the optimal $P_{1}^*,P_{2}^*$ for sorted pairing scheme can be derived as
\begin{equation}
\begin{array}{c}
  P_{1}^*=\frac{P_{\max}}{2}+\frac{H_1G_1-H_2G_2}{2H_1G_1H_2G_2},\ P_{2}^*=\frac{P_{\max}}{2}-\frac{H_1G_1-H_2G_2}{2H_1G_1H_2G_2}
\end{array},
\label{p1}
\end{equation}
and similarly, the optimal $P_{1}^{'*},P_{2}^{'*}$ for non-sorted pairing scheme can be derived as
\begin{equation}
\begin{array}{c}
  P_{1}^{'*}=\frac{P_{\max}}{2}+\frac{H_1G_2-H_2G_1}{2H_1G_1H_2G_2},\ P_{2}^{'*}=\frac{P_{\max}}{2}-\frac{H_1G_2-H_2G_1}{2H_1G_1H_2G_2}
\end{array}.
\label{p1'}
\end{equation}

Note that since only equality constraint in (\ref{C2}) is considered, (\ref{p1}) and (\ref{p1'}) is valid only for \textbf{Case 1} ($0<P_{1}^*, P_{2}^*, P_{1}^{'*}, P_{2}^{'*}<P_{\max}$). Due to the inequality constraints in (\ref{C2}), we need to consider another two cases, that is, \textbf{Case 2} ($P_{1}^*=P_{\max}, P_{2}^*=0, P_{1}^{'*}=P_{\max}, P_{2}^{'*}=0$) and \textbf{Case 3} ($P_{1}^*=P_{\max}, P_{2}^*=0, 0< P_{1}^{'*}, P_{2}^{'*}< P_{\max}$). For the remaining case ($0< P_{1}^*, P_{2}^*< P_{\max}, P_{1}^{'*}=P_{\max}, P_{2}^{'*}=0$), it is easy to find that it cannot occur, because it can be derived that $P_{1}^*$ is necessarily larger than $P_{1}^{'*}$ according to our assumptions. Noting that here we only consider $P_{1}^{'*}>P_{2}^{'*}$, that is, $H_1G_2-H_2G_1>0$, the analysis for $P_{1}^{'*}\leq P_{2}^{'*}$ is similar. Then we can derive and prove (\ref{prove2}) for the three cases, respectively. 

\textbf{Case 1}: For this case, using (\ref{p1}) and (\ref{p1'}), we can obtain
\[
\begin{array}{l}
  H_1G_1P_1^*+H_2G_2P_2^*+H_1G_1H_2G_2P_1^*P_2^*-(H_1G_2P_1^{'*}+H_2G_1P_2^{'*}+H_1G_2H_2G_1P_1^{'*}P_2^{'*})\\
  =\frac{(H_1-H_2)(G_1-G_2)}{2}P_{\max}+\frac{(H_1^2-H_2^2)(G_1^2-G_2^2)}{4H_1G_1H_2G_2}\\
  >0,
\end{array}
\]
where inequality is obtained from our assumption $H_1>H_2$ and the obtained result $G_1>G_2$.

\textbf{Case 2}: For this case, we can derive
\[
\begin{array}{l}
  H_1G_1P_1^*+H_2G_2P_2^*+H_1G_1H_2G_2P_1^*P_2^*-(H_1G_2P_1^{'*}+H_2G_1P_2^{'*}+H_1G_2H_2G_1P_1^{'*}P_2^{'*})\\
  =H_1P_{\max}(G_1-G2)\\
  >0,
\end{array}
\]
where inequality is obtained since $G_1>G_2$.

\textbf{Case 3}: For this case, using (\ref{p1'}), we can obtain
\begin{equation}
\begin{array}{l}
  H_1G_1P_1^*+H_2G_2P_2^*+H_1G_1H_2G_2P_1^*P_2^*-(H_1G_2P_1^{'*}+H_2G_1P_2^{'*}+H_1G_2H_2G_1P_1^{'*}P_2^{'*})\\
  =H_1G_1P_{\max}-(\frac{H_1G_2+H_2G_1}{2}P_{\max}+\frac{H_1G_2H_2G_1}{4}P_{\max}^2+\frac{(H_1G_2-H_2G_1)^2}{4H_1G_2H_2G_1}).
\end{array}
\label{cs3}
\end{equation}

For this case, it is found that the total power $P_{\max}$ satisfies $\frac{H_1G_2-H_2G_1}{H_1G_1H_2G_2}<P_{\max}\leq\frac{H_1G_1-H_2G_2}{H_1G_1H_2G_2}$. In this interval, we can prove (\ref{cs3}) is a monotonically increasing function of $P_{\max}$. So we substitute the lower bound of the interval $\frac{H_1G_2-H_2G_1}{H_1G_1H_2G_2}$ into (\ref{cs3}), and then we can derive
\[
\begin{array}{l}
  H_1G_1P_1^*+H_2G_2P_2^*+H_1G_1H_2G_2P_1^*P_2^*-(H_1G_2P_1^{'*}+H_2G_1P_2^{'*}+H_1G_2H_2G_1P_1^{'*}P_2^{'*})\\
  =\frac{2H1(H1G2-H2G1)(G1-G2)}{2H_1G_1H_2G_2}\\
  >0,
\end{array}
\]
where inequality is obtained from the aforemention condition $H_1G_2-H_2G_1>0$ and $G_1>G_2$.

Since (\ref{cs3}) is a monotonically increasing function of $P_{\max}$ in the whole interval, (\ref{cs3}) is always more than 0.

In summary, it is proved that in all cases, (\ref{prove1}) always holds. Therefore, for two-subcarrier case ($N = 2$), Theorem 1 is proved.

\emph{b)} Multi-subcarrier case ($N>2$):  The two-subcarrier case can be generalized to the multi-subcarrier case $N>2$. A proof by contradiction is used to prove Theorem 1 for the case $N>2$. For an $N$-subcarrier relay system with $N>2$, suppose the optimal pairing does not follow the sorted pairing rule of Theorem 1, so there are at least two pairs of incoming and outgoing subcarriers that are mismatched according to their channel gains. Without loss of generality, it is assumed there are two pairs ($i_1,j_1$) and ($i_2,j_2$) satisfying $\left| {h_{i_1}} \right|^2>\left| {h_{i_2}} \right|^2, \left| {g_{j_1}} \right|^2<\left| {g_{j_2}} \right|^2$. Using the result for $N=2$, it is found that pairing subcarrier $i_1$ with subcarrier $j_2$ and subcarrier $i_2$ with subcarrier $j_1$ can achieve a higher rate than the nonsorted pairings. Hence, by using this new pairing while maintaining the other subcarrier pairs invariant, the total achievable rate can be increased. This contradicts our assumption on the optimality of a nonsorted pairing scheme.




\ifCLASSOPTIONcaptionsoff
  \newpage
\fi

\newpage

\begin{figure*}[t]
\centering
\includegraphics[width=0.9\textwidth]{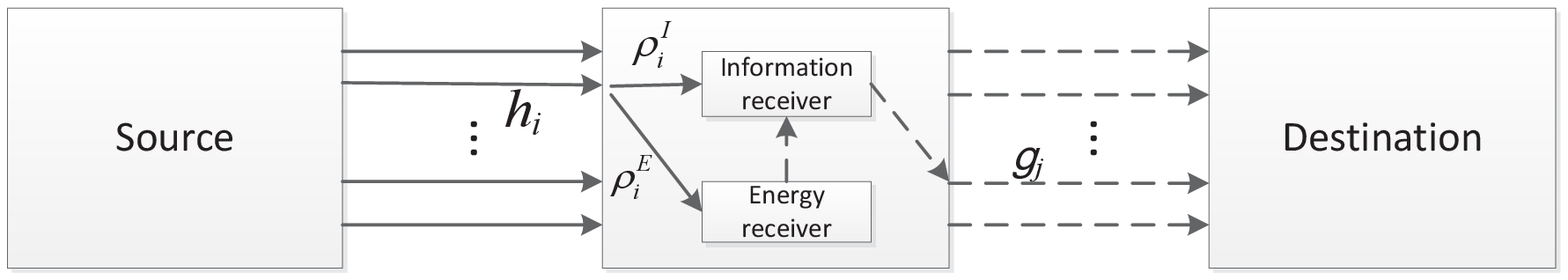}
\caption{Two-hop OFDM relay system, where the middle node represents the relay and it performs information decoding (ID) and energy harvesting (EH) by using PS scheme.}
\label{systemmodel}
\end{figure*}

\begin{figure*}[t]
\centering
\includegraphics[width=0.7\textwidth]{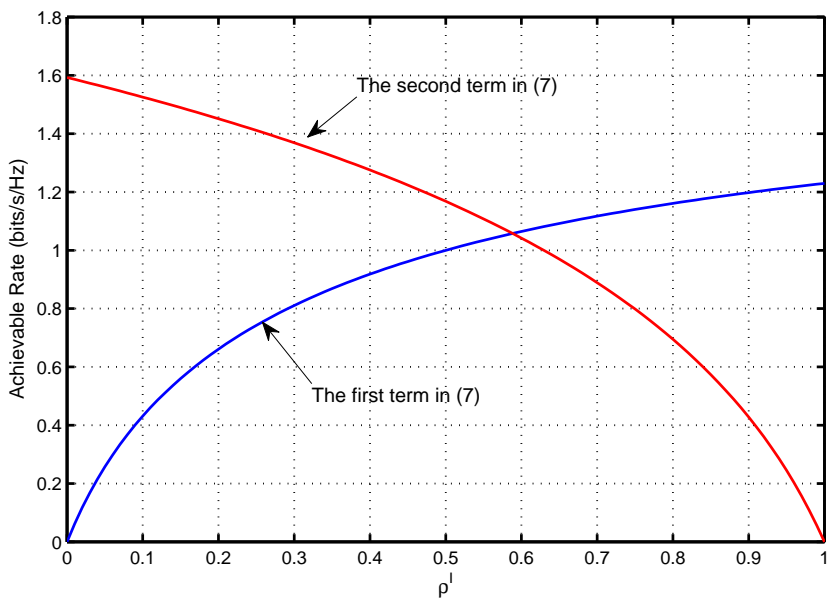}
\caption{A plot of the typical curves of two terms in (\ref{eq2}), the blue curve represents the first term and the red curve represents the second term for $\left| {{h}} \right|^2=\left| {{g}} \right|^2=0.9, \sigma_d^2=\sigma_{r,a}^2=\sigma_{r,b}^2=$1dBm, $\eta=1,P=$10dBm.}
\label{curves}
\end{figure*}

\begin{figure*}[t]
\centering
\includegraphics[width=0.7\textwidth]{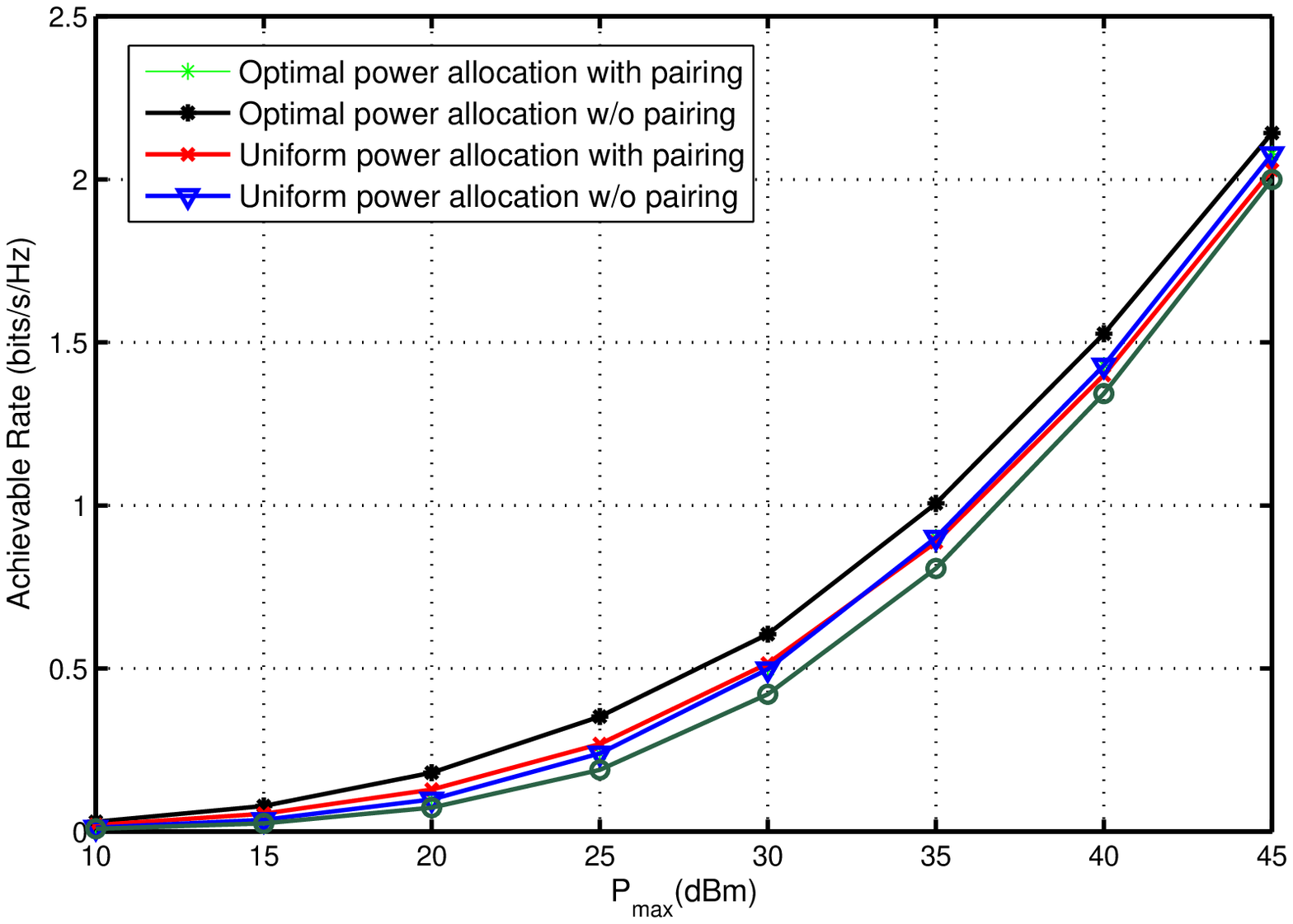}
\caption{Achievable transmission rate versus $P_{\max}$ with $d_0=1$ m and the relay located at the middle point between the source and destination.}
\label{Fig3}
\end{figure*}

\begin{figure*}[t]
\centering
\includegraphics[width=0.7\textwidth]{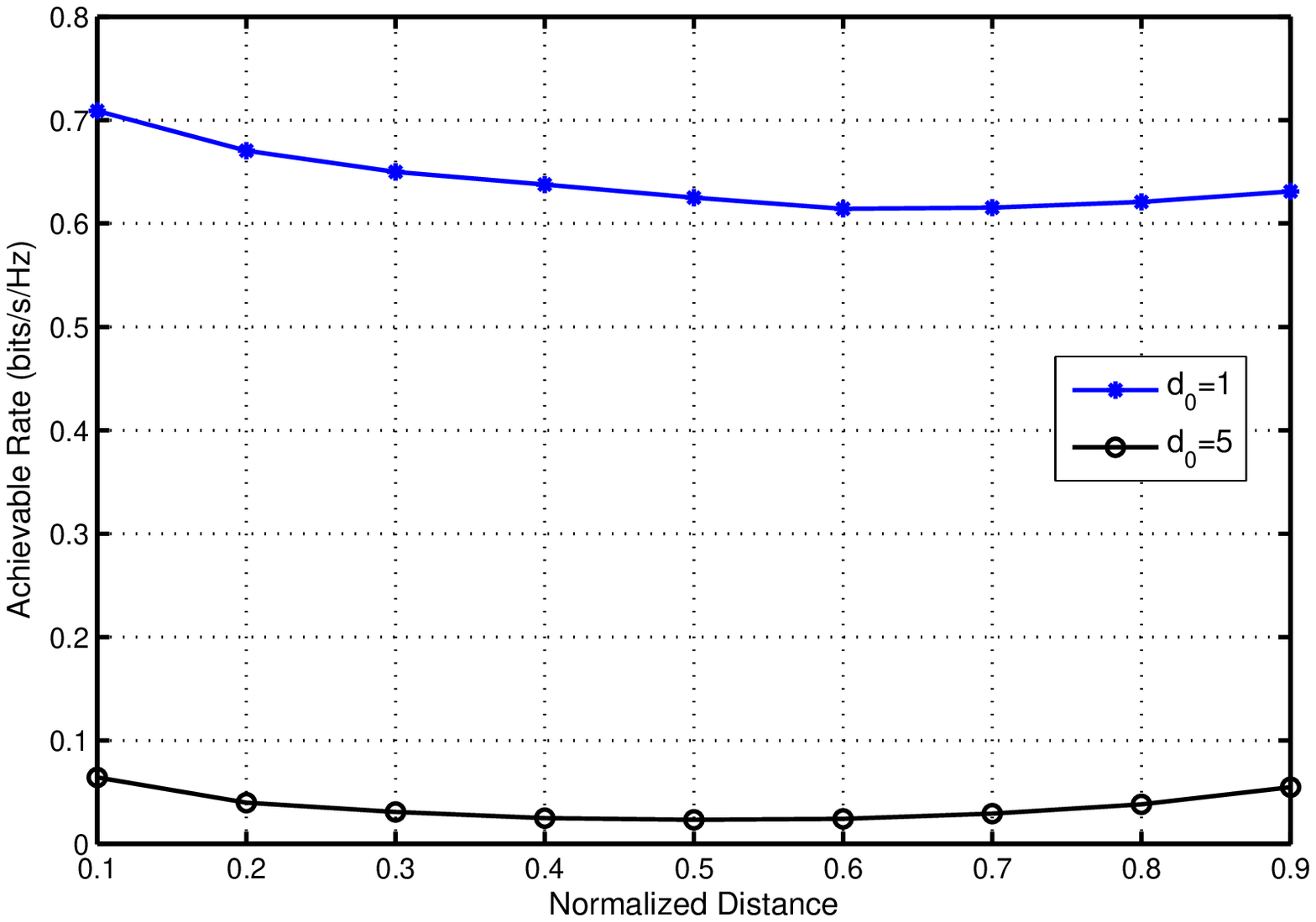}
\caption{Achievable transmission rate versus relay location with $P_{\max}$=30 dBm.}
\label{Fig4}
\end{figure*}

\begin{figure*}[t]
\centering
\includegraphics[width=0.7\textwidth]{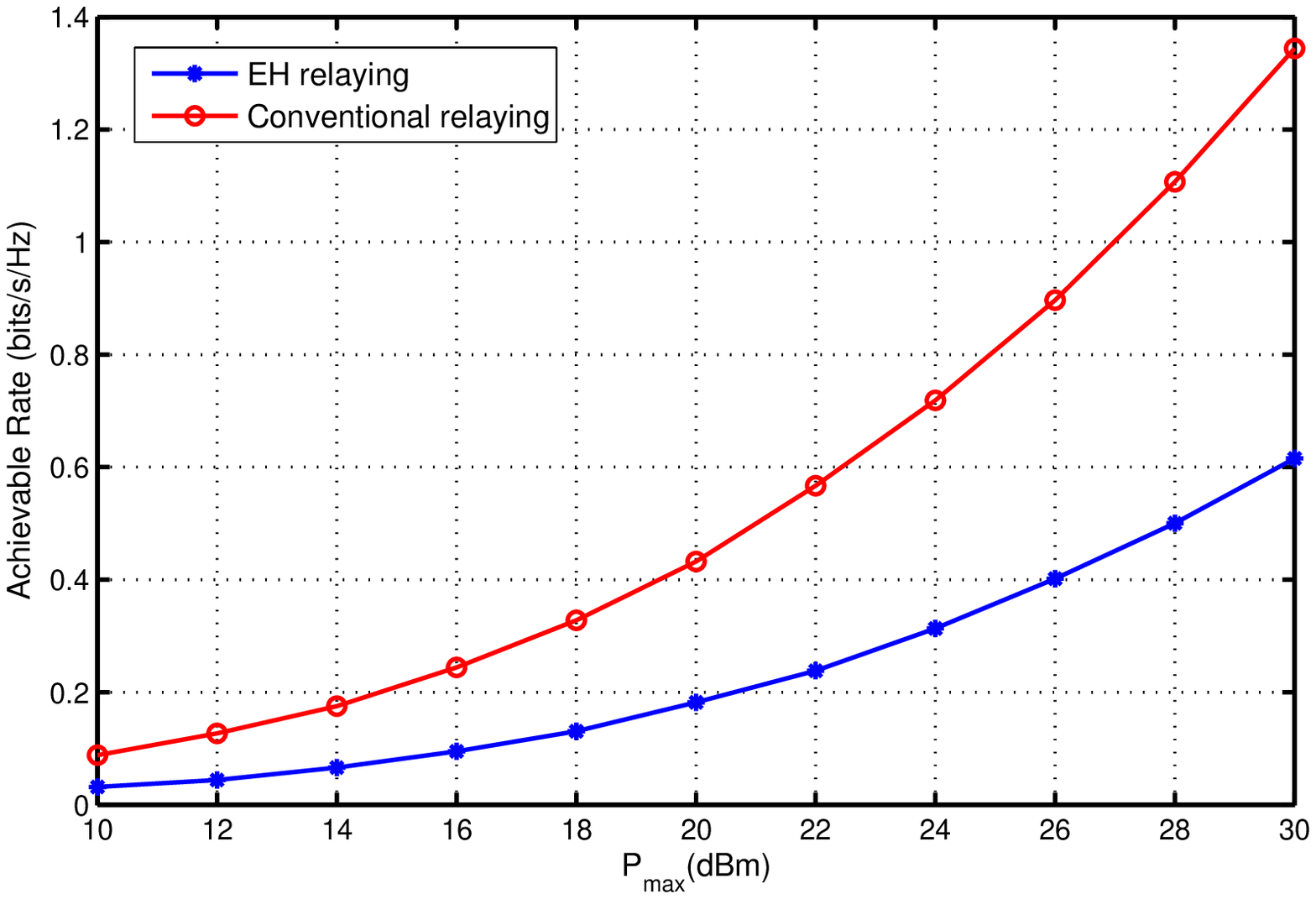}
\caption{Achievable transmission rate versus $P_{\max}$ with $d_0=1$ m and the relay located at the middle point between the source and destination.}
\label{Fig5}
\end{figure*}

\begin{figure*}[t]
\centering
\includegraphics[width=0.7\textwidth]{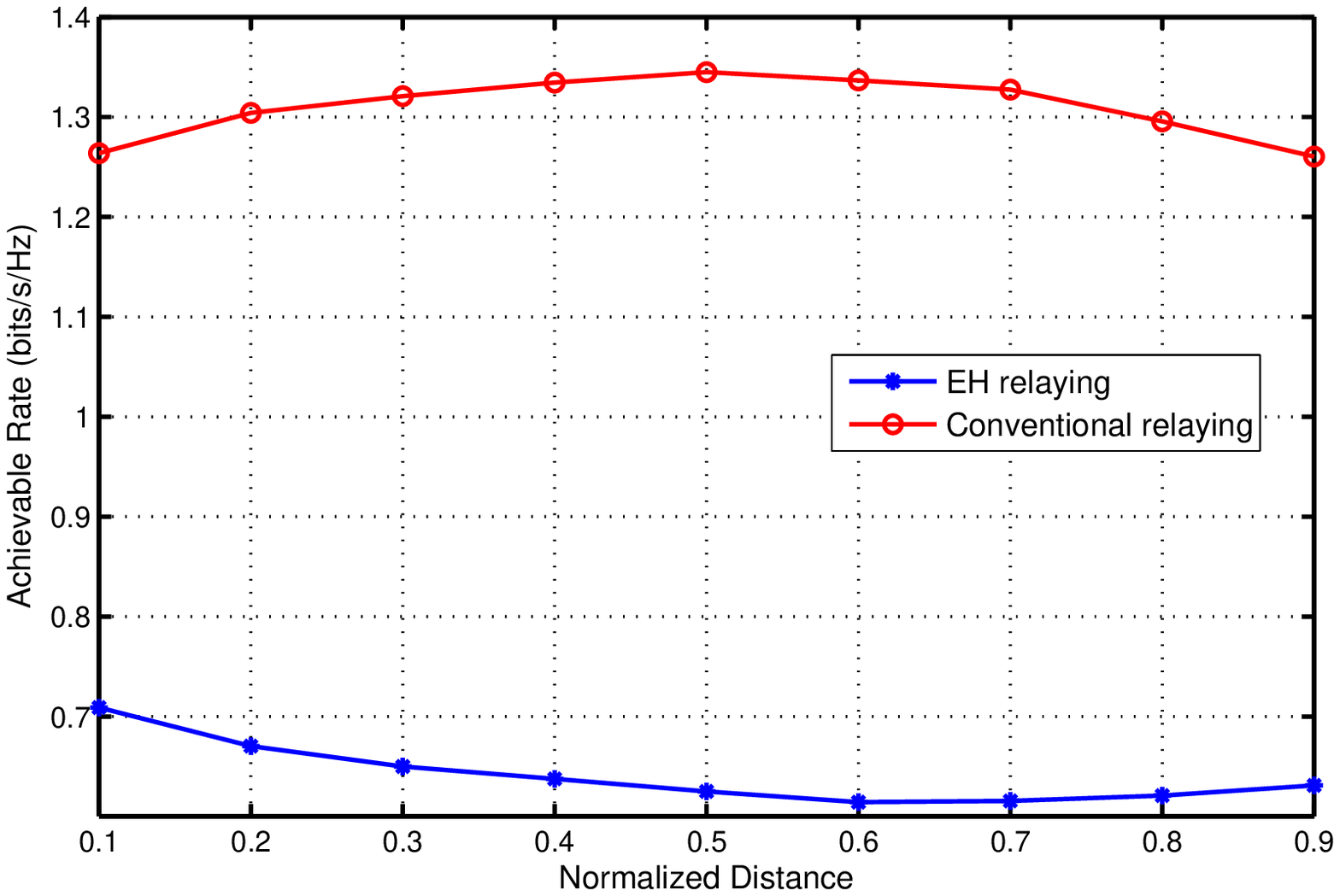}
\caption{Achievable transmission rate versus relay location with $P_{\max}$=30 dBm, $d_0=1$. }
\label{Fig6}
\end{figure*}




\end{document}